\documentclass[aps,prb,twocolumn,superscriptaddress,floatfix,showpacs]{revtex4}
\usepackage{graphicx}
\usepackage{amsmath}
\usepackage{amssymb}
\usepackage{bm}
\bibstyle{apsrev.bib}

\begin{document}
\title{Disorder-induced phases in the $S=1$ antiferromagnetic 
Heisenberg chain
}
\author{P\'eter Lajk\'o}
\affiliation{Department of Physics, Kuwait University, 
P.O. Box 5969, Safat, Kuwait 13060}
\author{Enrico Carlon}
\affiliation{Interdisciplinary Research Institute c/o IEMN, Cit\'e
Scientifique Bo\^ite Postale 60069, F-59652 Villeneuve d'Ascq, France}
\author{Heiko Rieger}
\affiliation{Theoretische Physik, Universit\"at des Saarlandes,
D-66041 Saarbr\"ucken, Germany}
\author{Ferenc Igl\'oi}
\affiliation{Research Institute for Solid State Physics and Optics, 
H-1525 Budapest, P.O. Box 49, Hungary}
\affiliation{Institute of Theoretical Physics, Szeged University,
H-6720 Szeged, Hungary}
\date{\today}

\begin{abstract}
  
  We use extensive density matrix renormalization group (DMRG) calculations to explore the phase diagram of
  the random $S=1$ antiferromagnetic Heisenberg chain with a power-law
  distribution of the exchange couplings.  We use open chains and
  monitor the lowest gaps, the end-to-end correlation function and the
  string order parameter. For this distribution, at weak disorder the
  system is in the gapless Haldane phase with a disorder dependent
  dynamical exponent, $z$, and $z=1$ signals the border between the
  nonsingular and singular regions of the local susceptibility. For
  strong enough disorder, which approximately corresponds to a
  uniform distribution, a transition into the random singlet phase is
  detected, at which the string order parameter as well as the average
  end-to-end correlation function are vanishing and at the same time
  the dynamical exponent is divergent. Singularities of physical
  quantities are found to be somewhat different in the random singlet phase
  and in the critical point.

\end{abstract}


\maketitle

\newcommand{\bc}{\begin{center}}
\newcommand{\ec}{\end{center}}
\newcommand{\be}{\begin{equation}}
\newcommand{\ee}{\end{equation}}
\newcommand{\ba}{\begin{array}}
\newcommand{\ea}{\end{array}}
\newcommand{\beqn}{\begin{eqnarray}}
\newcommand{\eeqn}{\end{eqnarray}}

\section{Introduction}

The $S=1$ spin antiferromagnetic Heisenberg chain has received much
attention, both experimentally and theoretically, since
Haldane\cite{haldane}  conjectured that its low-energy properties
are qualitatively different from that of the exactly solved $S=1/2$
model. The $S=1$ chain (together with all other integer spin chains) has
a finite gap in the excitation spectrum and hidden topological order
in the ground state, which is characterized by the string correlation
function.\cite{denN89} On the other hand, the bulk spin-spin correlations of the
model are short ranged, having a finite correlation length,
$\xi$. In an open chain of length $L$,
there are spin $S=1/2$ degrees of freedom at each edge and the
end-to-end correlations approach a finite value in an exponential
fashion, having the same characteristic length scale, $\xi$, as bulk
correlations.\cite{sorensenaffleck94}

Quenched disorder, which is realized by random couplings, also has different
effects for $S=1/2$ and $S=1$. In the former case any amount of disorder is
enough to drive the system into a new type of fixed point, \cite{fisherxx} whereas for the
$S=1$ chain, weak disorder is irrelevant and the properties of the weakly
random chain are the same as that of the pure one.\cite{hyman-dimer} For stronger disorder, however,
the low-energy properties of the system are changed and detailed analytical and
numerical investigations were devoted to clarify the properties of the new random fixed
points. 

The analytical studies of the random chain are made by variants of the strong
disorder renormalization group
(RG) method, which has been introduced for the $S=1/2$ chain by Ma, Dasgupta,
and Hu \cite{mdh} and
has been analyzed in great details by Fisher.\cite{fisherxx} This strong disorder RG
method has been used afterwards for a large variety of random quantum
and classical systems, (for a review, see Ref. \onlinecite{review}). For the $S=1$ chain, extensions of
the original Ma-Dasgupta rules are necessary\cite{Hyma97,monthus} to describe the disorder induced phases
in the system, which include a gapless Haldane (GH) phase, for intermediate disorder, and a
random singlet (RS) phase, for stronger disorder.

Numerical studies of the random $S=1$ chain have been made by exact
diagonalization\cite{nishiyama} of
the density matrix renormalization\cite{Hida}
(DMRG), by quantum Monte Carlo (QMC) methods,\cite{todo,bergkvist} and by numerical implementations of the strong
disorder RG method. \cite{saguia02} Despite considerable numerical effort, several aspects of the
low-energy properties of the random $S=1$ chain are still unclear and some numerical
results are conflicting.
In the numerical calculations mainly boxlike  distribution of disorder is considered,
which, as noted in Ref. \onlinecite{comment}, represent only a limited strength of randomness.
In numerical RG studies both the GH and the RS phases are identified; however, the transition
point between these phases is rather approximate.
In DMRG calculation (see also Ref. \onlinecite{nishiyama}) Hida \cite{Hida} has identified only
the GH phase, and conjectured that the
RS phase is not accessible for any finite strength  of disorder. In a comment to Hida's work,\cite{Hida}
Yang and Hyman\cite{comment} have predicted the appearance of the RS phase for some type of
power-law distribution of the disorder.
Another numerical work by QMC simulations \cite{todo} has shown the existence of the RS phase even for
the boxlike distribution and these results are confirmed by independent QMC simulations.\cite{bergkvist}
In the QMC calculations, some properties of the RS phase are verified (cf. scaling relation between length
and time, decay of the string correlation function), but results about
the spin-spin correlation function are different from the RG predictions. At the critical point no
numerical estimates are available to check analytical RG predictions. We note on recent studies
of Griffiths effects\cite{griffiths} in the system with enforced dimerization\cite{Daml02,arakawa}
and related work on the random $S=3/2$ and higher spin chains\cite{Refa02,clri04,segu03}.

In this paper, our aim is to study the low-energy properties of the random $S=1$ chain by the
DMRG method.\cite{DMRGbook} 
The features of our study are the following: (i) We consider a more general (power-law) distribution
of disorder, which allows us to enter more deeply into the RS phase, thus to obtain convincing evidence
of its existence. (ii) We calculate a different physical quantity, the end-to-end correlation function, which
carries important information about the phases of the system. The average end-to-end correlation
function has a finite limiting value in the GH phase and vanishes in the RS phase. Furthermore,
in the GH phase from the low-value tail of its distribution, independent estimates about the dynamical
exponent are obtained. (iii) We try to perform a comparative analysis between the properties of the
system at the critical point and in the RS phase and to check the available RG predictions.

The structure of our paper is the following. The model, the basic ingredients of the strong
disorder RG methods, and the conjectured phases are given in Sec. \ref{Sec:2}. Results of our
DMRG studies are presented in Sec. \ref{Sec:3} and discussed in Sec. \ref{Sec:4}.

\section{The model and the strong disorder RG results}
\label{Sec:2}
\subsection{Model}
We consider the spin $S=1$
random antiferromagnetic Heisenberg chain with the Hamiltonian
\begin{equation}
H = \sum_i J_{i} \vec{S}_i \cdot \vec{S}_{i+1},
\label{hamilton}
\end{equation}
where the $J_i > 0$ are independent and identically distributed random variables.
Here, we use the following power-law distribution
\be
p_\delta (J) = \delta^{-1} J^{-1+1/\delta}
\ \ \ \ \ \ \ \ \ \
{\rm for} \ 0 \leq J \leq 1,
\label{distribut}
\ee
where $\delta^2 ={\rm var}[ \ln J]$ measures the strength of disorder. In previous numerical
work, a boxlike distribution was used,
\be
P_W(J)=\left\{
\begin{aligned}
&1/W& &{\rm for}\quad 1-W/2<J<1+W/2\\
&0& & {\rm otherwise,}
\end{aligned}
\right.
\label{box}
\ee
in which the strength of disorder grows with $W$. Note that the possible maximal value, $W=2$, corresponds
to the uniform distribution, which can be obtained from  Eq. (\ref{distribut}) with $\delta=1$ and having
a prefactor, $1/2$, and a range $0 \le J \le 2$. Consequently, the
power-law distribution for $\delta>1$ represents a disorder, which is stronger than any boxlike
disorder.

The low-energy behavior of the system of size, $L$, is encoded in the distribution of the
lowest gap, $\Delta$, denoted by $P_L(\Delta)$. We note that for an open chain the first gap
corresponds to the localized edge states; therefore, one should study the second (not localized)
gap. The average spin-spin correlation function is denoted by
\be
C(i,j)=[\langle S^{z}_i S^{z}_j \rangle]_{\rm av},
\label{cij}
\ee
where $[\cdots ]_{\rm av}$ stands for averaging over quenched
disorder. For bulk correlations with $|j-i| \ll i,j=O(L)$, we have $C(i,j)=C_b(|j-i|)$,
whereas for end-to-end correlations, $C(1,L) \equiv C_1(L)$.
The string correlation function of the model is defined by\cite{denN89}
\begin{equation}
O^z(r)=- \langle S_l^z \exp\left[ i \pi \left( S_{l+1}^z+S_{l+2}^z+\dots+S_{l+r-1}^z \right)\right] S_{l+r}^z \rangle\;,
\label{string}
\end{equation}
and its large $r$ limiting value is the string order parameter. For several quantities it turned
out useful to consider the average of its inverse. More
precisely, for a physical observable, $f$, we denote by $f^{\rm iv}$ the following quantity:
\be
f^{\rm iv}=\frac{1}{[ f^{-1} ]_{\rm av}}\;,
\label{inverse}
\ee
what we shall call as inverse average.

\subsection{Weak disorder limit--Haldane phase}
\label{limiting}

In absence of randomness ($J_{i} = J$) the spectrum is gapped,\cite{haldane} and bulk
spin-spin correlations are short ranged, $C_b(r) \sim \exp(-r/\xi)$ with
$\xi=6.03$\cite{sorensenaffleck94}. On the contrary, end-to-end spin-spin correlations
and the string correlation function have a finite limiting value.
For weak disorder, when the distribution of $J$ is sufficiently narrow,
the Haldane gap is robust and the system stays in
the Haldane phase.\cite{hyman-dimer} The border of the Haldane phase can be estimated by
noting that the Haldane gap is robust against enforced dimerization,\cite{ah87}
when even and odd couplings are different, so that  
\begin{equation}
J_i=J(1+D (-1)^i)\exp(\delta \zeta_i)\;,
\label{J_S}
\end{equation}
where $\zeta_i$ are random numbers of mean zero and variance unity.
The pure system ($\delta=0$) for $D<0.25$ stays in the Haldane
phase\cite{singh} and at the phase transition point the coupling at an
odd bond, $J_o$, and that at an even bond, $J_e$, are related as $J_o
= 0.6 J_e$. We expect that in the presence of disorder the Haldane gap
stays finite, if the maximum ($J_{max}$) and the minimum ($J_{min}$)
values of the couplings satisfy $J_{min}/J_{max} > 0.6$. From this
argument we obtain for the border of the Haldane phase for the box
distribution $W_G \approx 0.5$. On the other hand, for the power-law
distribution in Eq. (\ref{distribut}) $J_{min}=0$; therefore, for any
$\delta>0$ the Haldane phase is expected to be destroyed.

\subsection{Strong disorder limit--RG approach}

For strong disorder the low-energy properties of the system are
explored by variants of the strong disorder RG approach.  In the
standard Ma-Dasgupta--type RG approach, the couplings of the random
antiferromagnetic Heisenberg chain are put in descending order and the
largest coupling defines the energy scale, $\Omega$, in the system.
During renormalization the pair of spins with the largest coupling,
say $J_i=\Omega$, are replaced by a singlet and decimated out. At the
same time a new coupling is generated between the spins at the two
sides of the singlet, which is given in a perturbation calculation as
\be
\tilde{J}=\frac{4}{3} \frac{J_{i-1} J_{i+1}}{J_i}\;.
\label{J_pert}
\ee
As noticed by Boechat, Saguia, and Continentino\cite{saguia96} for weak disorder some of
the generated new couplings can be larger than the energy scale,
$\Omega$. Therefore, the standard strong disorder RG approach works
only for strong enough disorder and describes only the RS phase of
the system.

To cure this problem, different types of RG approaches are proposed.
Monthus  \it et al. \rm \cite{monthus}  suggested to replace the pair of spin
$S=1$ connected by the strongest bond by a pair of $S=1/2$. In this
case the renormalized system consists of a set of spin $S=1$ and
$S=1/2$ degrees of freedom, having both antiferromagnetic and
ferromagnetic couplings. The renormalized couplings, which are
calculated perturbatively, are all smaller than $\Omega$. This RG
approach, during which no spin larger than $S=1$ is generated, can be
used to describe both the gapless Haldane and the RS phases and
provides precise numerical estimates about the critical exponents.

In another modified RG approach, Saguia \it et al. \rm \cite{saguia02} use the
standard perturbative approach in Eq.~(\ref{J_pert}), provided
$max(J_{i-1},J_{i+1})<3\Omega/4$.  Otherwise the triplet of spins with
couplings, $max(J_{i-1},J_{i+1})$, and $\Omega$ is replaced by a single
spin. Also in this method the variation of the energy scale is
monotonic: the generated two new couplings are both smaller than
$\Omega$.

Recently, a variant of the strong disorder RG method was proposed by one of
us,\cite{Peter} in
which the pair of spin with the strongest coupling is decimated out,
but--and this is a feature of our current method--the new coupling
between the remaining spins is calculated nonperturbatively. The four
spins with couplings $J_{i-1},~J_{i}$, and $J_{i+1}$ are replaced by
two spins and during decimation the lowest gap in the two systems
remains the same. It is easy to see that the rule we use is somewhat similar
to the approach by Saguia \it et al. \rm  \cite{saguia02} However, this method has no
discontinuity in the approximation, which could be important in the
vicinity of the critical point, where a crossover takes place between
the different decimation regimes.

\subsection{Disorder-induced phases}

Based on a modified strong disorder RG approach\cite{Hyma97,monthus,saguia02} and different
numerical calculations,\cite{Hida,todo,bergkvist} the following scenario of the
phase transition in the model is conjectured with increasing strength
of disorder.

\subsubsection{Gapless Haldane phase}

For sufficiently strong disorder ($\delta>\delta_G$ or $W > W_G$), the
gap in the Haldane phase is closed and one arrives to the gapless
Haldane phase. As we have argued in Sec. \ref{limiting}, $\delta_G=0$ and
$W_G \approx 0.5$. The GH phase is a quantum Griffiths
phase,\cite{griffiths} in which the correlation length, $\xi(\delta)$,
is finite, whereas the typical time scale, $t_r \sim \Delta^{-1}$, is
divergent. Relation between the size of the system, $L$, and the
smallest gap is given by
\be
\Delta \sim L^{-z}\;,
\label{z'}
\ee
where $z$ is the disorder induced dynamical exponent. The
distribution of the lowest gap is given by
\begin{equation}
P_L(\Delta){\rm d} \Delta= L^{-z}\tilde{P}\left[\frac{\Delta}{L^z}\right] {\rm d} \Delta\;,
\label{eq:scaling-conv}
\end{equation}
and $\tilde{P}(x) \sim x^{-1+1/z}$ for small $x$, so that from the
low-energy tail $z$ can be calculated.  Similarly the distribution of
the end-to-end correlation function has a vanishing tail, which
behaves as\cite{ijl01} $P(C_1) \sim C_1^{-1+1/z}$, which gives an independent
way to calculate the
dynamical exponent.  Some thermodynamical
quantities such as the local susceptibility, $\chi$, and the specific
heat, $c_v$, are singular at low temperature,\cite{review}
\be
\chi(T) \sim T^{-1+1/z},\quad c_v(T) \sim T^{1/z}\;.
\label{T}
\ee
The limit of divergence of $\chi(T)$ is signaled by $z=1$, and the corresponding disorder
is denoted by $\delta_1$ ($W_1$). The separation of the two parts of the GH phase with
$z<1$ and $z>1$ can
be located by considering the inverse average of the gap, $\Delta^{\rm iv}$, and
the inverse average of the end-to-end correlation function, $C_1^{\rm iv}$. In the
nonsingular region, $z<1$, both $\Delta^{\rm iv}$ and $C_1^{\rm iv}$ are finite,
whereas in the singular region, $z>1$, both are vanishing.

To see this, we consider the inverse average of the gap
\be
\Delta^{\rm iv} \sim \left[ \int_{\Delta_{min}}^{\Delta_{max}} \Delta^{-2+1/z} {\rm d} \Delta\right]^{-1}
\sim \frac{z-1}{\Delta_{min}^{-1+1/z}-\Delta_{max}^{-1+1/z}}\;,
\label{inv}
\ee
which indeed tends to zero, if $z>1$ and $\Delta_{min} \to 0$. On the other hand for the vanishing of
the average gap one needs $\Delta_{max} \to 0$. One can use a similar reasoning for the end-to-end
correlation function, for which the upper limit of the distribution, $C_1^{max}>0$, thus
$[C_1]_{\rm iv}>0$, in the whole region, $\delta<\delta_1$.

In a static sense, the gapless Haldane phase is noncritical: the average end-to-end correlation
function, as well as the string order parameter, is finite in the complete GH phase.

\subsubsection{Critical point}

Increasing the strength of disorder over a critical value ($\delta_C$ or $W_C$), the system arrives
at the random singlet phase. As the critical strength of disorder is approached, the correlation
length diverges: $\xi \sim (\delta_C-\delta)^{-\nu}$, with $\nu=(1+\sqrt{13})/2$ and the string
order parameter vanishes as\cite{Hyma97,monthus} $O^z(\delta) \sim
(\delta_C-\delta)^{2 \beta}$, with $\beta=[2(3-\sqrt 5)/
(\sqrt13 -1)]$. At the critical point the string order-parameter decays
algebraically, $O^z(r) \sim r^{-\eta^{st}}$, with $\eta^{st}=2\beta/\nu$.
The end-to-end correlation function goes to zero algebraically, too,
$C_1(L) \sim L^{-\eta_1}$, similarly to the
bulk spin-spin correlation function, $C_b(r) \sim r^{-\eta}$. Here, however, there are no theoretical conjectures
about the exponents $\eta_1$ and $\eta$.
The relation at the critical point between the correlation length and the relaxation time is strongly
anisotropic,
\be
\ln \, t_r \sim \xi^{\psi}\;,
\label{psi}
\ee
with $\psi=1/3$; thus, the dynamical exponent, $z$, is formally infinity. This type of infinite disorder
scaling is seen in the distribution of the gaps, which is given by
\begin{equation}
P_L(\Delta){\rm d} \Delta=
L^{-\psi}\tilde{P}\left[\frac{\ln \, {\Delta}}{L^\psi}\right] {\rm d} \, \ln \, \Delta\;.
\label{eq:scaling-psi}
\end{equation}
In the space of variables, dimerization ($D$) and disorder ($\delta$),
the critical point of the system represents a multicritical point in which three
Griffiths phases with different symmetry meet.\cite{Daml02} The corresponding
exponents follow by permutation symmetry and the calculation can be generalized
for higher values of $S$.\cite{Daml02}

\subsubsection{Random singlet phase}

For a disorder $\delta>\delta_C$ ($W>W_C$), the low-energy behavior of
the system is controlled by an infinite disorder fixed point and the
system is in the RS phase. The RS phase is a critical phase, both
$\xi$ and $t_r$ are divergent, and its properties are assumed to be
identical to the RS phase of the random $S=1/2$ chain. This latter
system is studied in great detail by Fisher\cite{fisherxx} with the
asymptotically exact strong disorder RG method and these results have
been confronted with detailed numerical
investigations.\cite{henelius,ijr00,stolze} Here we repeat that
in the RS phase there is infinite disorder scaling, so that relations in
Eqs. (\ref{psi}) and (\ref{eq:scaling-psi}) are valid with an exponent,
$\psi=1/2$.  The RS phase is instable against enforced dimerization,
as given in Eq. (\ref{J_S}), and the correlation length behaves as
$\xi(D) \sim D^{-\nu}$, with $\nu=2$. In the RS phase the bulk and
end-to-end correlation functions decay algebraically.  In Table
\ref{table1} we collected the conjectured values of the critical
exponents both in the random singlet phase and at the critical point and
compared these values with the estimates obtained in this paper.

\begin{table}
\caption{Theoretical predictions for the critical exponents in the random singlet (RS) phase and at
the critical point (CP). Values obtained in this paper are given in square brackets.
\label{table1}}
 \begin{tabular}{|c|c|c|c|c|c|c|}  \hline
   & $\eta^{st}$ & $\beta$  & $\nu$     & $\psi$   & $~\eta~$ & $~\eta_1~$\\ \hline
RS & $0.382$[0.41(4)] & - &      2     &     1/2 [0.45(5)]    &    2  &  1 [0.86(6)]   \\
CP & $0.509[0.39(3)]$ & $0.586$  & $2.30$     &    1/3 [0.35(5)]  & - & - [0.69(5)]     \\  \hline

  \end{tabular}
  \end{table}

\subsection{Summary of the existing numerical results}

In previous numerical studies the box distribution in Eq. (\ref{box})
has been used. In Table \ref{table2} we present the estimates of the
borders of the different phases obtained by different numerical
methods, such as by numerical implementation of the strong disorder
RG, by DMRG, and by QMC. We note that for the power-law disorder in
Eq. (\ref{distribut}) the critical disorder is estimated\cite{comment}
as $\delta_C \approx 1.5$. Using variants of the strong disorder RG
method,\cite{monthus,saguia02} the calculated critical exponents in the RS phase--within
numerical precision--correspond to the predicted, analytical
values. To reach the asymptotic region,
however, one often needs to treat very long chains of length
$L=10^4-10^6$, see Ref. \onlinecite{monthus}. The DMRG and QMC investigations have led
to different conclusions for the strongest box disorder, with $W=2$.
No RS phase is found by DMRG, \cite{Hida} whereas by QMC infinite
disorder scaling is detected.\cite{todo,bergkvist} The average string correlation function
was shown to decay algebraically with\cite{bergkvist} $\eta_{st}=0.378(6)$, close to
the theoretical result in Table \ref{table1}. On the other hand, the
average spin-spin correlation function was found to have an exponent,\cite{bergkvist}
$\eta=1$, which greatly differs from the theoretical value of
$\eta=2$.

\begin{table}
\caption{Numerical estimates for the borders of the different phases
of the random $S=1$ chain with boxlike disorder, see Eq. (\ref{box}). The different phases are
defined by
$W<W_G$, Haldane phase; $W_G < W < W_1$, GH phase with nondivergent
local susceptibility; $W_1 < W < W_C$, GH phase with divergent
local susceptibility; and $W>W_C$, RS phase.
\label{table2}}
 \begin{tabular}{|c|c|c|c|}  \hline
   & $W_G$   & $W_1$     & $W_C$  \\ \hline
RG[\onlinecite{monthus}] & & & 1.48   \\\hline
RG[\onlinecite{saguia02}] & &0.76 & 2.   \\\hline
DMRG[\onlinecite{Hida}] & &1.8 & no   \\\hline
QMC[\onlinecite{todo}] & 1.37 & 1.7 & 1.8   \\\hline
QMC[\onlinecite{bergkvist}] &  &  & $<2$   \\  \hline

  \end{tabular}
  \end{table}

\section{Numerical investigations}
\label{Sec:3}
\subsection{The DMRG method}

Most of our numerical results are based on DMRG calculations. In this
case we used open chains up to length $L=64$, for weak disorder and up to
$L=32$ for strong disorder and calculated the
lowest two gaps, the end-to-end correlation function and the string
order parameter. This latter quantity is calculated for open chains between
points $i=L/4$ and $j=3L/4$. Note that for an open chain the first gap is related to
the surface degrees of freedom and goes to zero exponentially with $L$. The
characteristic bulk excitations are given by the second gap and we studied
this quantity. In the numerical calculation we have retained up to $m=180$
states in the DMRG and checked that convergence of the numerical results is reached. We used
the power-law distribution of disorder in Eq. \!\!(\ref{distribut}) in the canonical ensemble, i.e., there was
no constraint to the value of the sum of the odd and even couplings. In this way there is a
nonzero residual dimerization, which could be the source of some error for small systems.
However, using the microcanonical ensemble, in which the sum of the odd couplings is the same
as that of the even couplings, could leave to different finite-size exponents for the end-to-end
correlation function, which is known for the random transverse-field Ising chain.\cite{bigpaper,young,cecile}
We have calculated typically 10 000 independent disorder realizations in each case.

\subsection{Gapless Haldane phase}

\subsubsection{Nonsingular region: $z<1$ }

\begin{figure}
\includegraphics [width=0.7 \linewidth]{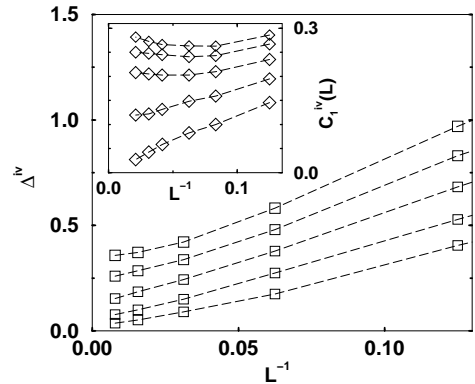}
\caption{The inverse average gap, $\Delta^{\rm iv}$, as a function of the
inverse size of the system for different strengths of
disorder, $\delta=0.1, 0.2, 0.3, 0.4$, and $0.5$ from the top to the bottom, respectively. $\Delta^{\rm iv}$
seems to vanish around $\delta \approx 0.5$. Inset: the inverse average end-to-end
correlation function as a function of
inverse size, with the same values of disorder, as in the main panel. Note that for weak disorder
the size dependence of $C_1^{\rm iv}$ is nonmonotonic, which is due to a finite correlation length
in the system.
\label{fig:avgap}
}
\end{figure}

We have calculated the distribution of the (second) gap and determined
its inverse average, $\Delta^{\rm iv}$,
which is presented in Fig. \ref{fig:avgap} as a function of the inverse size, $L^{-1}$, for
different values of $\delta$. The limiting value as $L \to \infty$ is monotonously decreasing
with $\delta$ and $\Delta^{\rm iv}$ seems to approach zero at a limiting disorder $\delta_1\approx 0.45-0.5$.
Similar conclusion is obtained from the behavior of the inverse average of the end-to-end
correlation function, $C^{\rm iv}_1(L)$, which is shown in the inset of Fig. \ref{fig:avgap}.

\begin{figure}
\includegraphics [width=0.7 \linewidth]{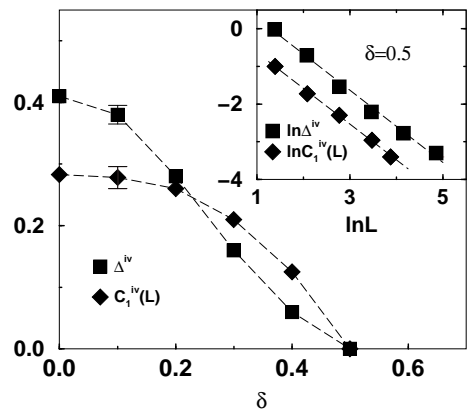}
\caption{Extrapolated values of the inverse average gap and the inverse average
  end-to-end correlation function as a function of the strength of
  disorder. At $\delta=0$ we obtain estimates for the nonrandom
  model, $\Delta=0.4105(3)$ and $C_1=0.283(1)$. Inset: Size dependence of the
  inverse average gap and the inverse average end-to-end correlation
  function in a log-log plot at the boundary of the gapless Haldane
  phase. Both lines have an approximate slope, $z=1$, denoted by broken lines.
The typical value of the error is indicated; otherwise, the error is smaller than
the size of the symbol.
\label{fig:extrG}
}
\end{figure}

Note, however, that the average end-to-end correlation function, as shown in
Fig. \ref{fig:CP} is finite
at $\delta_1$.
The extrapolated values of $\Delta^{\rm iv}$ and $C^{\rm iv}_1(L)$ are shown
in Fig. \ref{fig:extrG}. Close to $\delta_1$, both curves are compatible with an approximately linear
variation with $\delta_1-\delta$. At the boundary point, $\delta_1=\delta$, the size dependences
of $\Delta^{\rm iv}$ and $C^{\rm iv}_1(L)$ are shown in the inset of Fig. \ref{fig:extrG}. Both are
linear in $L^{-1}$, in accordance with the criterion that at $\delta_1$
 the disorder induced dynamical exponent is $z(\delta_1) = 1$.

\subsubsection{Singular region: $z>1$ }

\begin{figure}
\includegraphics [width=0.7 \linewidth]{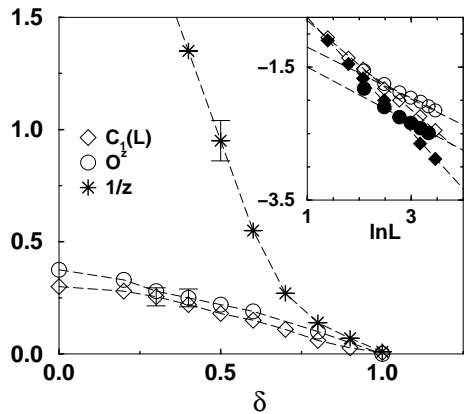}
\caption{The average string order parameter, $O^z$, the average end-to-end
  correlation function, $C_1(L)$ and the disorder induced dynamical
  exponent, $1/z$, as a function of the strength of disorder. Inset:
  finite-size dependence of the string order parameter and the average
  end-to-end correlation function at the critical point ($\delta_c=1$,
  open symbols) and in the RS phase ($\delta_c=1.5$, full symbols) in
  a log-log plot. The slope of the broken lines representing the
  critical exponents are $\circ, ~0.39 \pm0.03$[0.509];
  $\bullet, ~0.41\pm0.04[0.382]$; $\lozenge, ~0.69\pm0.05$; and
  $\blacklozenge, ~0.86\pm0.06[1.0]$ where in the brackets we presented
  the theoretical RG results; see Table \ref{table1}. Typical values
  of the error are indicated; otherwise, the error is smaller than the
  size of the symbol.
\label{fig:CP}
}
\end{figure}

\begin{figure}
\includegraphics [width=0.7 \linewidth]{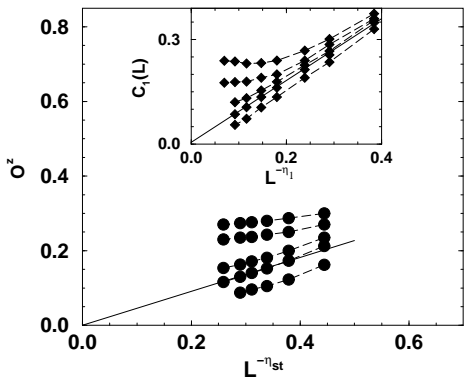}
\caption{The average string orderparameter as a function of $L^{-\eta_{st}}$,
with $\eta_{st}=0.39$ as obtained in the inset of Fig. \ref{fig:CP},
for disorder, $\delta=0.3, \, 0.5, \,  0.8, \, 1.0,$ and $1.5$ from the top to the
  bottom.  Inset: the average end-to-end correlation function as a function of $L^{-\eta_1}$,
$\eta_1=0.69$ being the critical decay exponent, for the same values of disorder as in the main panel. 
Solid straight lines over the $\delta=1.0$ points are guide to the eyes.
\label{fig:GH}
}
\end{figure}

We have calculated the average string order parameter and the average
end-to-end correlation function for different sizes $L$. Also we have
determined the disorder induced dynamical exponent, $z$, which is
deduced from the low-energy tail of the gap distribution [see
Eq. (\ref{eq:scaling-conv})]. The extrapolated values of $O^z$ and
$C_1$, as well as $1/z$ are plotted in Fig. \ref{fig:CP} for different
strengths of disorder. All these three quantities tend to zero around the
same limiting value of disorder and
the border of the Griffiths phase, i.e., the location
of the critical point can be determined as $\delta_c=1.0(1)$.
In the extrapolation procedure we have made use of the finite-size dependence
of $O^z(L) \sim L^{-\eta_{st}}$ and $C_1(L)\sim L^{-\eta_{1}}$ at the critical point,
which is shown in the inset of Fig. \ref{fig:CP}. For weaker disorder, $\delta<\delta_c$,
$O^z$ tends to a finite limiting value, which is illustrated in Fig. \ref{fig:GH} using a scale
$L^{-\eta_{st}}$. A similar conclusion is obtained for the average end-to-end correlation
function, which is presented in the inset of Fig. \ref{fig:GH}.

\subsection{Critical point and the RS phase}

Our aim with the numerical investigations in this subsection is twofold: first, to
check the properties of the RS phase, thus to present numerical evidences, and second,
and this is numerically more demanding, to try to discriminate between the properties in
RS phase and at the critical point. We start to analyze the finite-size dependence
of the average string order parameter and that of the
average end-to-end correlation function, which is shown in the inset of Fig. \ref{fig:CP}
at two values of the disorder, $\delta=1$ and $\delta=1.5$. The first value should be
close to the critical point (see Fig. \ref{fig:CP}); however, there is certainl;y some uncertainty,
see the values of $W_G$ in Table \ref{table2}. The second value of disorder, $\delta=1.5$,
should be deeply in the RS phase; however, see the RG estimates in Ref. \onlinecite{comment}.

At $\delta=1.5$ the decay of the average
string order parameter, as well as that of the end-to-end correlation function is
algebraic, and the decay exponents of both quantities are in agreement with the theoretical
prediction in the RS phase, as given in Table \ref{table1}. For the average end-to-end correlation
function we have somewhat less accuracy, which could be due to similar crossover effects as
noted for the bulk spin-spin correlation function in Ref. \onlinecite{bergkvist}.
The same type of analysis of the results at $\delta=1$ give somewhat
different results. The decay of
the average end-to-end correlation function is characterized by an exponent,
$\eta_1=0.69$, which differs considerably from the value in the RS phase. On the
other hand, the decay exponent of the average string order parameter, within the error of the
calculation, agrees with the value found in the RS phase. We note that
at the same disorder in the QMC simulation\cite{bergkvist} also the exponent in
the RS phase is observed. One possible explanation is that $\delta=1$ is already in the RS phase
and therefore we find the corresponding exponent. Anyway, even at the critical point one expects strong
crossover effects due to the vicinity of the RS fixed point, so that probably much larger
systems are needed to observe the true asymptotic behavior.

\begin{figure}
\includegraphics [width=0.7 \linewidth]{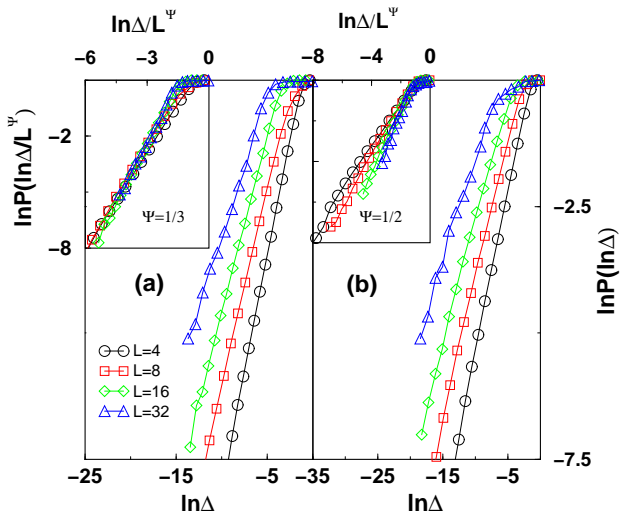}
\vspace{5mm}
\caption{(Color online) Distribution of the gaps in finite systems at the critical point, $\delta=1$ (a);
and in the RS phase, $\delta=1.4$ (b). In the insets scaling collapse with
Eq. (\ref{eq:scaling-psi})
is shown with $\psi=1/3$ (a) and $\psi=1/2$ (b).
\label{fig:SC}
}
\end{figure}

Finally, we compare in Fig. \ref{fig:SC}
the distribution of the gaps at the critical point (a) and in the RS phase (b). For
both cases the distribution is broadened with increasing $L$, which is a clear signal of infinite
disorder scaling. Indeed, one can obtain a good scaling collapse using the form
in Eq. \!(\ref{eq:scaling-psi}). In the insets we have illustrated this type of behavior
by using the theoretical predictions, $\psi=1/3$ at the
critical point and $\psi=1/2$ in the RS phase, respectively. The estimated exponents obtained from
the optimal scaling collapse are shown in Table \ref{table1}.

\section{Conclusion}
\label{Sec:4}

The random antiferromagnetic $S=1$ chain is a paradigm of disorder
induced phase transition phenomena (see also Ref. \onlinecite{Carl01}) for
which detailed strong disorder RG predictions are available. These
predictions, however, have only been partially verified by numerical
calculations and even the numerical results are somewhat conflicting.
In this paper we have used extensive DMRG calculations with the aim to
clarify the low-energy properties of the system with varying strengths
of disorder. The sizes of the systems we used in the calculation are
comparable with those in previous DMRG studies; \cite{Hida} however, we
used a power-law distribution of the couplings in
Eq. \!(\ref{distribut}), which can be more random, than the box
distribution in Eq. \!(\ref{box}) used previously. We have also
calculated a quantity, the end-to-end spin-spin correlation
function, which can be used to locate the borders of the different
phases in the system and to obtain an independent estimate of the
dynamical exponent. Our calculations gave further numerical support of
the phase diagram predicted by the strong disorder RG method and our
results are basically in agreement with the scenario of disorder
induced phase transitions. In the RS phase we made calculations far
from the critical point, which is not possible with boxlike
distribution of couplings as given in Eq. (\ref{box}) and obtained
estimates for the critical exponents which are compatible with the RG
predictions. Our results at the critical point are less conclusive,
which is probably due to crossover effects and the inaccurate
location of this point. For the critical exponent of the end-to-end
correlation function, $\eta_1$, and that of the gap scaling, $\psi$,
numerical estimates at the critical point are clearly different from
that in the RS phase, which are in accordance with the RG results. On
the other hand, for the average string correlation function our
numerical results are in conflict with the RG prediction. We believe
that at this point much larger finite systems are necessary to obtain
a precise numerical estimate and thus to be able to test the results
of RG predictions.

We close our paper by mentioning that the present day numerical possibilities to
explore the properties of the random $S=1$ chain seem to be exhausted, as far as
DMRG or QMC methods are considered. Some independent and probably more accurate
results can be expected, however, by the numerical application of different variants
of the strong disorder RG method, in particular in the vicinity of the
critical point and in the RS phase. Results obtained in this direction will
be published in the future.\cite{Peter}

\centerline{\bf ACKNOWLEDGMENT}
This work has been supported by Kuwait University Research Grant No. [SP 09/02].

\end{document}